\documentclass[aps,twocolumn]{revtex4-1}
\usepackage{amsmath,amssymb,bm,graphicx,hyperref}
\allowdisplaybreaks[1]

\renewcommand\d{\partial}
\newcommand\grad{\bm{\nabla}}
\newcommand\+{\dagger}
\newcommand\up{\uparrow}
\newcommand\down{\downarrow}
\newcommand\eps{\epsilon}
\newcommand\A{\bm{A}}
\newcommand\x{{\bm{x}}}
\newcommand\z{{\bm{z}}}
\newcommand\eff{\mathrm{eff}}
\newcommand\GL{\mathrm{GL}}
\newcommand\MF{\mathrm{MF}}
\newcommand\Tr{\mathrm{Tr}}

\begin{document}

\title{Two-dimensional Fermi gas in antiparallel magnetic fields}

\author{Takaaki Anzai}
\author{Yusuke Nishida}
\affiliation{Department of Physics, Tokyo Institute of Technology,
Ookayama, Meguro, Tokyo 152-8551, Japan}

\date{January 2017}

\begin{abstract}
We study a two-dimensional Fermi gas with an attractive interaction subjected to synthetic magnetic fields assumed to be mutually antiparallel for two different spin components.
By employing the mean-field approximation, we find that its phase diagram at zero temperature consists of pair superfluid and quantum spin Hall insulator phases and closely resembles that of the Bose-Hubbard model.
The resulting two phases are separated by a second-order quantum phase transition classified into the universality class of either the dilute Bose gas or the $XY$ model.
We also show that the pairing gap can be enhanced significantly by the antiparallel magnetic fields as a consequence of magnetic catalysis, which may facilitate the realization of the pair superfluid in two dimensions by ultracold atom experiments.
\end{abstract}

\maketitle

\section{Introduction}
After the first realization of the Bose-Einstein condensation (BEC) in ultracold atomic gases~\cite{Anderson:1995,Davis:1995,Bradley:1995}, overwhelming experimental progress has been made to allow us to control system parameters at will~\cite{Bloch:2008}.
For example, the interaction between atoms can be controlled by varying a magnetic field through a Feshbach resonance~\cite{Inouye:1998,Courteille:1998}.
This technique applied to two-component Fermi gases led to the realization of a crossover to a Bardeen-Cooper-Schrieffer (BCS) superfluid of Cooper pairs from a BEC of tightly bound molecules~\cite{Regal:2004,Zwierlein:2004}.
Furthermore, the dimensionality can be controlled by confining atoms with an optical lattice generated by two counterpropagating laser beams~\cite{Bloch:2005}.
Therefore, the BCS-BEC crossover in two dimensions (2D) has also come close to the reach of experimental investigation~\cite{Modugno:2003,Martiyanov:2010,Frohlich:2011,Dyke:2011,Feld:2011,Sommer:2012,Vogt:2012,Zhang:2012,Koschorreck:2012,Baur:2012,Frohlich:2012,Makhalov:2014,Ong:2015,Ries:2015,Murthy:2015,Fenech:2016,Boettcher:2016,Cheng:2016}, which may provide important insights into layered superconductors~\cite{Randeria:1989,Randeria:1990}.

More recently, enormous research efforts have been devoted to develop experimental techniques to create synthetic magnetic fields~\cite{Dalibard:2011,Goldman:2014}.
One approach is to couple internal states of atoms by laser beams so that neutral atoms behave like charged particles in a magnetic field~\cite{Lin:2009}, which opened up a new avenue toward the realization of quantum Hall physics with ultracold atoms.
This approach was further extended to create ``antiparallel'' magnetic fields, which act on two different spin components of atoms with the same magnitude but in opposite directions~\cite{Beeler:2013}.
While they were implemented for a two-component Bose gas of rubidium atoms to observe a spin Hall effect, the same technique is in principle applicable to two-component Fermi gases.
There have also been proposals to create antiparallel magnetic fields by inducing laser-assisted tunneling in a tilted optical lattice~\cite{Aidelsburger:2013,Kennedy:2013}, aiming at the realization of the quantum spin Hall (QSH) effect~~\cite{Kane:2005a,Kane:2005b,Bernevig:2006}.

Motivated by these experimental abilities to control the interaction, dimensionality, and magnetic fields, we study a 2D Fermi gas with an attractive interaction between two spin components in antiparallel magnetic fields.
While the attractive interaction generally favors the Cooper pairing, the antiparallel magnetic fields lead to Landau-level formation with opposite chiralities for different spin components of fermions.
How do they compete or cooperate to give rise to interesting physics?
This is the subject to be elucidated in this Rapid Communication.
Besides its own importance, our system may also be viewed as a simulator of analogous phenomena in other fields, such as exciton condensation and chiral condensation in a magnetic field, where two particles forming a pair have opposite charges and thus experience opposite Lorentz forces~\cite{Eisenstein:2004,Miransky:2015}.
For related theoretical works on 2D Bose and 3D Fermi gases as well as in an optical lattice, see Refs.~\cite{Furukawa:2014,Feng:2015,Iskin:2015}.

In what follows, we set $\hbar=1$ and denote the magnetic length and the cyclotron frequency by $\ell_B\equiv1/\sqrt{B}$ and $\omega_B\equiv B/m$, respectively.
We also use shorthand notations, $(x)\equiv(\tau,\x)$, $\int\!dx\equiv\int_0^\beta\!d\tau\!\iint_0^L\!d^2\x$, $(k)\equiv(\omega_n,k_x,l)$, $\sum_k\equiv\sum_{\omega_n}\sum_{k_x}\sum_{l=0}^\infty$, and $\delta_{kk'}\equiv\delta_{\omega_n\omega'_n}\delta_{k_xk'_x}\delta_{ll'}$, where $\omega_n\equiv2\pi(n+1/2)/\beta$ and $k_x\equiv2\pi n/L$ are the Matsubara frequency and the wave number, respectively, and $l=0,1,2,\dots$ labels Landau levels.
The inverse temperature $\beta$ and the linear size of the system $L$ are formally kept finite in Sec.~\ref{sec:formulation}, while the zero temperature and thermodynamic limits $\beta,L\to\infty$ are taken at the end.

\section{Functional integral formulation\label{sec:formulation}}
Let us start with spin-$1/2$ fermions in 2D subjected to spin-dependent vector potentials, which are described by the following imaginary-time action:
\begin{align}
S &= \sum_{\sigma=\up,\down}\int\!dx\,\phi_\sigma^*(x)
\left[\d_\tau+\frac{[-i\grad+\A_\sigma(\x)]^2}{2m}-\mu\right]\phi_\sigma(x) \notag\\
&\quad - g\int\!dx\,\phi_\up^*(x)\phi_\down^*(x)\phi_\down(x)\phi_\up(x).
\end{align}
Here $m$ and $\mu$ are the mass and chemical potential common to both spin components of fermions and the coupling constant $g>0$ is assumed to be attractive.
We also choose the vector potentials as
\begin{align}
\A_\up(x) = -\A_\down(x) = -By\hat\x,
\end{align}
so that different spin components experience antiparallel magnetic fields with the magnitude $B>0$; $\grad\times\A_\up(x)=-\grad\times\A_\down(x)=B\hat\z$.
Note that this particular way of introducing the magnetic fields preserves the time-reversal symmetry as well as the spin conservation. 

To facilitate our theoretical analysis, it is convenient to employ the Hubbard-Stratonovich transformation~\cite{Altland-Simons},
\begin{align}
& S' = \int\!dx\,\frac{|\Delta(x)|^2}{g} - \int\!dx\,\Phi^\+(x) \notag\\
&\times
\begin{bmatrix}
-\d_\tau{-}\frac{[-i\grad+\A_\up(\x)]^2}{2m}{+}\mu & \Delta(x) \\
\Delta^*(x) & -\d_\tau{+}\frac{[-i\grad-\A_\down(\x)]^2}{2m}{-}\mu
\end{bmatrix}
\Phi(x),
\end{align}
so that the interaction term in the action is decoupled at the expense of introducing the pair field $\Delta(x)$.
We then expand the Nambu-Gor'kov spinor $\Phi(x)\equiv[\phi_\up(x),\phi_\down^*(x)]^T$ over the eigenfunctions of the single-particle Hamiltonian.
Here the eigenfunction in the Landau gauge is
\begin{align}
\chi_k(x) \equiv \frac{e^{-i\omega_n\tau+ik_xx}}{\sqrt{\beta L}}\,F_l(y-k_x\ell_B^2)
\end{align}
with
\begin{align}
F_l(y) \equiv \frac{e^{-(y/\ell_B)^2/2}}{\sqrt{2^ll!\pi^{1/2}\ell_B}}\,
H_l\!\left(\frac{y}{\ell_B}\right)
\end{align}
being the $l$th eigenfunction of the harmonic oscillator, which solves the Schr\"odinger equation $[(-i\grad-By\hat\x)^2/(2m)]\chi_k(x)=\eps_l\chi_k(x)$ with the single-particle energy provided by $\eps_l\equiv(l+1/2)\omega_B$~\cite{Landau-Lifshitz}.
By substituting the resulting expansion $\Phi(x)=\sum_k\chi_k(x)\tilde\Phi(k)$, the action is now expressed as
\begin{align}
S' = \int\!dx\,\frac{|\Delta(x)|^2}{g}
- \sum_k\sum_{k'}\tilde\Phi^\+(k)G^{-1}(k,k')\tilde\Phi(k'),
\end{align}
where the inverse Nambu-Gor'kov propagator is defined by
\begin{align}
G^{-1}(k,k') \equiv
\begin{bmatrix}
(i\omega_n-\xi_l)\delta_{kk'} & \tilde\Delta(k,k') \\
\tilde\Delta^*(k',k) & (i\omega_n+\xi_l)\delta_{kk'}
\end{bmatrix}
\end{align}
with $\xi_l\equiv\eps_l-\mu$ and $\tilde\Delta(k,k')\equiv\int\!dx\,\chi_k^*(x)\Delta(x)\chi_{k'}(x)$.

Finally, by integrating out the fermion fields, we obtain the effective action written in terms of the pair field:
\begin{align}\label{eq:effective}
S_\eff = \int\!dx\,\frac{|\Delta(x)|^2}{g} - \Tr\ln[G^{-1}(k,k')].
\end{align}
While this expression is formally exact with the understanding of the renormalization procedure discussed below, some approximation needs to be employed to proceed.
We first employ the mean-field approximation in Sec.~\ref{sec:mean-field} to investigate the phase diagram at zero temperature and then employ the Ginzburg-Landau expansion in Sec.~\ref{sec:universality} to elucidate the universality class of quantum phase transitions therein.

\section{Mean-field phase diagram at zero temperature\label{sec:mean-field}}
\subsection{Phase boundary}
To investigate the phase diagram at zero temperature, we employ the mean-field approximation.
By setting $\Delta(x)=\Delta_0>0$ and thus $\tilde\Delta(k,k')=\Delta_0\delta_{kk'}$, the thermodynamic potential $\Omega_\MF\equiv S_\eff/(\beta L^2)$ is found to be
\begin{align}
\Omega_\MF = \frac{\Delta_0^2}{g}
- \frac{m\omega_B}{2\pi}\sum_{l=0}^\infty(E_l-\xi_l)\theta(\Lambda-\eps_l),
\end{align}
where $E_l\equiv\sqrt{\xi_l^2+\Delta_0^2}$ is the quasiparticle energy and an energy cutoff $\Lambda$ is introduced because the second term turns out to be logarithmically divergent.
This logarithmic divergence should be canceled by the same form of divergence hidden in the coupling constant~\cite{Marini:1998}:
\begin{align}\label{eq:coupling}
\frac1g = \frac{m}{2\pi}\int_0^\Lambda\!\frac{d\epsilon}{2\eps+\eps_b}.
\end{align}
Here $\eps_b>0$ has the physical meaning of the binding energy of a two-body bound state in the vacuum without magnetic fields, which always exists for any $g>0$ in 2D and thus can be used to parametrize the attraction~\cite{Randeria:1989,Randeria:1990}.
By separating out the divergent piece from the second term, combining it with the first term, and then taking the limit of $\Lambda\to\infty$, the thermodynamic potential is made manifestly cutoff independent as
\begin{align}\label{eq:potential}
\Omega_\MF &= \frac{m\Delta_0^2}{4\pi}\left[\ln\!\left(\frac{2\omega_B}{\eps_b}\right)
+ \psi\!\left(\frac12-\frac\mu{\omega_B}\right)\right] \notag\\
&\quad - \frac{m\omega_B}{2\pi}\sum_{l=0}^\infty
\left(E_l-\xi_l-\frac{\Delta_0^2}{2\xi_l}\right),
\end{align}
where $\psi(z)$ is the digamma function.

The order parameter $\Delta_0$ is determined so as to minimize the thermodynamic potential (\ref{eq:potential}).
For any fixed chemical potential $\mu/\omega_B$ but not right at a Landau level (i.e., $\xi_l\neq0$ for all $l\in\mathbb{N}_0$), we find a second-order quantum phase transition from a normal state with $\Delta_0=0$ to a superfluid state with $\Delta_0>0$ by increasing the two-body binding energy $\eps_b/\omega_B$.
The resulting phase boundary is obtained by solving $\d^2\Omega_\MF/\d\Delta_0^2=0$ at $\Delta_0=0$ [cf.\ Eq.~(\ref{eq:gap}) below], which leads to
\begin{align}\label{eq:boundary}
\frac{\epsilon_b}{\omega_B}
= 2\exp\!\left[-\psi\!\left(\frac12-\frac\mu{\omega_B}\right)
+ 2\psi\!\left(\frac12-\frac\mu{\omega_B}+\nu\right)\right]
\end{align}
with $\nu\equiv\lfloor\mu/\omega_B+1/2\rfloor\theta(\mu)$ being the filling factor per spin.

\begin{figure}[t]
\includegraphics[width=0.9\columnwidth]{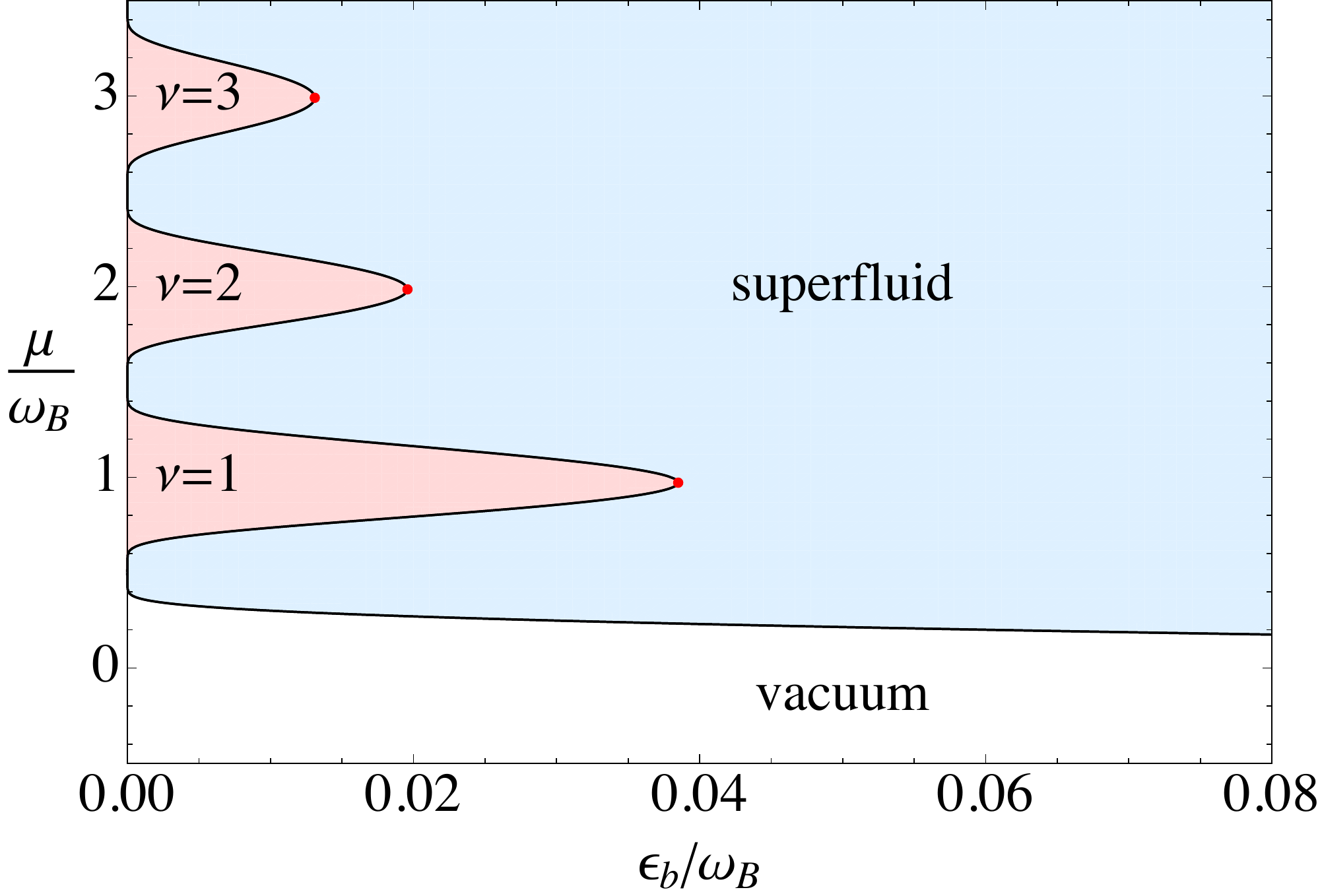}
\caption{\label{fig:phase_boundary}
Mean-field phase diagram at zero temperature in the plane of the two-body binding energy $\eps_b$ and the chemical potential $\mu$ in units of the cyclotron frequency $\omega_B$.
There are three types of phases corresponding to the vacuum, the pair superfluid, and the quantum spin Hall insulator labeled by its filling factor $\nu=1,2,3,\dots$.
They are separated by the second-order quantum phase transition located at Eq.~(\ref{eq:boundary}).}
\end{figure}

As one can see from the phase diagram depicted in Fig.~\ref{fig:phase_boundary}, the normal state with $\Delta_0=0$ is divided into different phases corresponding to different filling factors $\nu=0,1,2,\dots$.
The number density therein is provided by $n=(m\omega_B/\pi)\nu$ and thus the phase for $\nu=0$ is just the vacuum where no particles are present.
On the other hand, for $\nu>0$, each spin component of fermions fills $\nu$ Landau levels so that our system becomes the QSH insulator composed of two quantum Hall states with opposite chiralities for different spin components~\cite{Kane:2005a,Kane:2005b,Bernevig:2006}.
The rest of the phase diagram is occupied by the pair superfluid state with the nonzero pairing gap $\Delta_0>0$, whose behavior shall be investigated further.

\subsection{Pairing gap}
The BCS-BEC crossover in the superfluid phase is described by the number density equation,
\begin{align}\label{eq:number}
n = -\frac{\d\Omega_\MF}{\d\mu}
= \frac{m\omega_B}{2\pi}\sum_{l=0}^\infty\left(1-\frac{\xi_l}{E_l}\right),
\end{align}
together with the gap equation, $\d\Omega_\MF/\d\Delta_0=0$:
\begin{align}\label{eq:gap}
\ln\!\left(\frac{2\omega_B}{\eps_b}\right) + \psi\!\left(\frac12-\frac\mu{\omega_B}\right)
- \omega_B\sum_{l=0}^\infty\left(\frac1{E_l}-\frac1{\xi_l}\right) = 0.
\end{align}
In the strong-coupling limit $\epsilon_b\to\infty$, we have $-\mu\gg\Delta_0$ so that the number density equation reduces to $n\simeq(m/4\pi)\Delta_0^2/(-\mu)$, while the gap equation reduces to $\ln(-2\mu/\epsilon_b)+(\Delta_0/\mu)^2/4\simeq0$.
Therefore, we find
\begin{align}
\mu \to \eps_F - \frac{\epsilon_b}{2}
\qquad\text{and}\qquad
\Delta_0 \to \sqrt{2\eps_F\epsilon_b}
\end{align}
with $\eps_F\equiv\pi n/m$ being the Fermi energy, which coincide with the results for the 2D BCS-BEC crossover without magnetic fields~\cite{Randeria:1989,Randeria:1990}.
This is understandable because our system in the strong-coupling limit consists of tightly bound spin-singlet molecules, for which the antiparallel magnetic fields cancel out.

On the other hand, the superfluid phase can persist down to the weak-coupling limit $\epsilon_b\to0$ only when the chemical potential lies right at a Landau level, i.e., $\xi_l=0$ for some $l\in\mathbb{N}_0$.
The number density in this case reduces to $n\to(m\omega_B/\pi)(l+1/2)$, while the gap equation is solved by
\begin{align}\label{eq:weak-coupling}
\frac{\Delta_0}{\omega_B} \to \frac1{\ln(2\omega_B/\eps_b)-2\gamma-\psi(l+1)}.
\end{align}
Therefore, we find that the pairing gap in terms of the small coupling constant (\ref{eq:coupling}) is expressed as
\begin{align}
\frac{\Delta_0}{\omega_B} \to \frac{mg}{4\pi},
\end{align}
which exhibits the remarkable linear dependence in contrast to the usual exponential dependence without magnetic fields; $\Delta_0\propto e^{-2\pi/(mg)}$~\cite{Randeria:1989,Randeria:1990}.
This results from the divergent density of states at Landau levels and such an enhancement of dynamical symmetry breaking by magnetic fields is generally referred to as ``magnetic catalysis''~\cite{Miransky:2015}.

\begin{figure}[t]
\includegraphics[width=0.9\columnwidth]{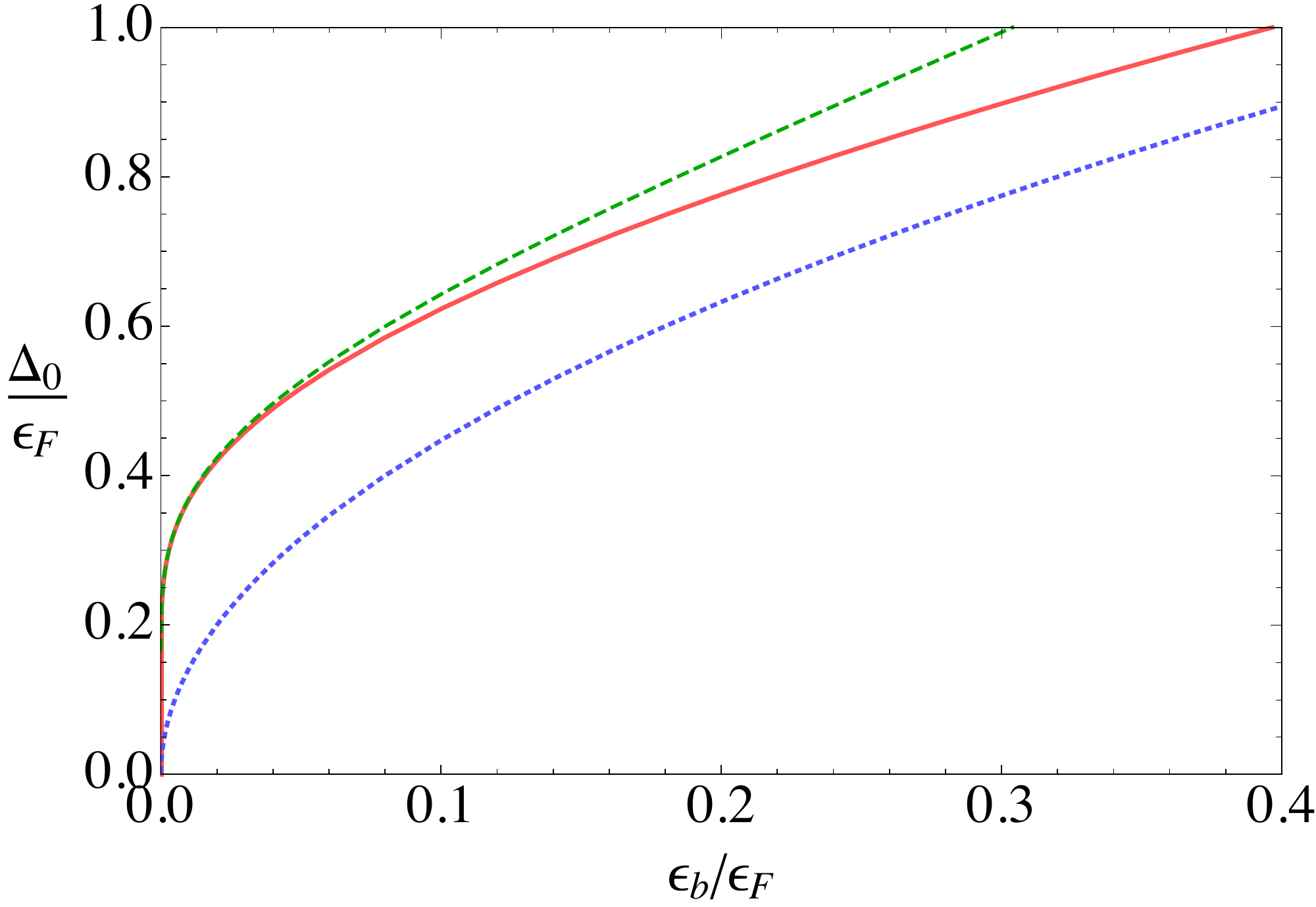}
\caption{\label{fig:pairing_gap}
Mean-field pairing gap $\Delta_0$ as a function of the two-body binding energy $\eps_b$ in units of the Fermi energy $\eps_F\equiv\pi n/m$.
The solid curve shows the result in the presence of the antiparallel magnetic fields with $\omega_B/\eps_F=2$, while the dotted curve shows $\Delta_0=\sqrt{2\eps_F\eps_b}$ in their absence.
Its asymptotic behavior in the weak-coupling limit [Eq.~(\ref{eq:weak-coupling})] is also indicated by the dashed curve.}
\end{figure}

The pairing gap beyond the weak-coupling limit is plotted in Fig.~\ref{fig:pairing_gap}, which is obtained by solving the coupled Eqs.~(\ref{eq:number}) and (\ref{eq:gap}).
Here one can see that the pairing gap is indeed enhanced significantly by the antiparallel magnetic fields, which may facilitate the realization of the pair superfluid in 2D by ultracold atom experiments.

\section{Universality class of quantum phase transitions\label{sec:universality}}
The phase diagram obtained in the previous section closely resembles that of the Bose-Hubbard model which consists of superfluid and Mott insulator phases~\cite{Fisher:1989}.
Here it was revealed that the quantum phase transition between them is classified into the universality class of either the dilute Bose gas or the $XY$ model.
This fact and the mutual resemblance motivate us to elucidate the universality class of the quantum phase transition in our system.

In the vicinity of the quantum phase transition, the effective action (\ref{eq:effective}) can be expanded with respect to the pair field $\Delta(x)$ assumed to be small and smooth.
By keeping terms up to the quartic order in $\Delta(x)$ and the quadratic order in derivatives, the Ginzburg-Landau action after some straightforward calculations is found to be
\begin{align}\label{eq:ginzburg-landau}
S_\GL &= \int\!dx\,\bigl[a_1\Delta^*(x)\d_\tau\Delta(x) + a_2|\d_\tau\Delta(x)|^2 \notag\\
&\quad + b_2|\grad\Delta(x)|^2 + c_2|\Delta(x)|^2 + c_4|\Delta(x)|^4\bigr]\,.
\end{align}
Here an unimportant constant term is dropped and the other coefficients are provided by
\begin{align}
a_1 &= -\frac{m}{8\pi\omega_B}\left[\psi'\!\left(\frac12-\frac\mu{\omega_B}\right)
- 2\psi'\!\left(\frac12-\frac\mu{\omega_B}+\nu\right)\right], \\
a_2 &= c_4 = \frac{m}{32\pi\omega_B^2}\left[\psi''\!\left(\frac12-\frac\mu{\omega_B}\right)
- 2\psi''\!\left(\frac12-\frac\mu{\omega_B}+\nu\right)\right], \\
b_2 &= \frac\mu{8\pi\omega_B^2}\,\biggl[2\psi\!\left(\frac12-\frac\mu{\omega_B}\right)
- \psi\!\left(-\frac\mu{\omega_B}\right) - \psi\!\left(1-\frac\mu{\omega_B}\right) \notag\\
& - 4\psi\!\left(\frac12{-}\frac\mu{\omega_B}{+}\nu\right)
+ 2\psi\!\left(-\frac\mu{\omega_B}{+}\nu\right)
+ 2\psi\!\left(1{-}\frac\mu{\omega_B}{+}\nu\right)\biggr]\,, \\
c_2 &= \frac{m}{4\pi}\left[\ln\!\left(\frac{2\omega_B}{\eps_b}\right)
- \psi\!\left(\frac12-\frac\mu{\omega_B}\right)
+ 2\psi\!\left(\frac12-\frac\mu{\omega_B}+\nu\right)\right].
\end{align}
While $a_2$, $b_2$, and $c_4$ are always positive, $c_2$ changes its sign when the phase boundary located at Eq.~(\ref{eq:boundary}) is crossed.
Furthermore, because we can find the relationship,
\begin{align}
a_1 = -\frac12\frac{\d c_2}{\d\mu},
\end{align}
$a_1$ vanishes at $\d c_2/\d\mu=0$ corresponding to the tip of each QSH phase marked by the red dot in Fig.~\ref{fig:phase_boundary}.

When $a_1=0$, the Ginzburg-Landau action (\ref{eq:ginzburg-landau}) is invariant under the exchange of $\Delta(x)\leftrightarrow\Delta^*(x)$, which signals the particle-hole symmetry emergent in the low-energy limit.
Accordingly, the equation of motion obeyed by the pair field, $\delta S_\GL/\delta\Delta^*(x)=0$, becomes the nonlinear Klein-Gordon equation, which is relativistic and Lorentz invariant.
Here the quantum phase transition turns out to be in the universality class of the $XY$ model~\cite{Sachdev}.
On the other hand, away from the tip of the QSH phase, the $a_2$ term in the action is negligible with respect to the nonvanishing $a_1$ term.
In this case, the equation of motion is the usual Gross-Pitaevskii equation and thus the quantum phase transition falls into the universality class of the dilute Bose gas~\cite{Sachdev}.

\section{Conclusion and outlook}
In this Rapid Communication, we studied competition and cooperation between the attractive interaction and the antiparallel magnetic fields in a 2D Fermi gas.
When the chemical potential does not match any Landau levels, the antiparallel magnetic fields compete with the Cooper pairing, i.e., our system becomes a QSH insulator and can be a pair superfluid only by a sufficient attraction (see Fig.~\ref{fig:phase_boundary}).
By employing the mean-field approximation at zero temperature, we found that these two phases are separated by a second-order quantum phase transition.
Its universality class turns out to be of the $XY$ model at the tip of each QSH phase and of the dilute Bose gas elsewhere, which closely resembles the phase diagram of the Bose-Hubbard model.

On the other hand, when the chemical potential matches some Landau level, the antiparallel magnetic fields in turn cooperate with the Cooper pairing, i.e., not only our system can be a pair superfluid by an infinitesimal attraction, but also the pairing gap is significantly enhanced (see Fig.~\ref{fig:pairing_gap}).
In particular, we showed that the usual exponential dependence on the small coupling constant is promoted to the remarkable linear dependence as a consequence of magnetic catalysis.
Although the role of fluctuations still needs to be understood, our finding here may facilitate the realization of the pair superfluid in 2D by ultracold atom experiments.

As for a future work, we plan to extend our study to finite temperature as well as to finite density imbalance between two different spin components of fermions.
In particular, it was revealed in the absence of magnetic fields that the Fulde-Ferrell-Larkin-Ovchinnikov (FFLO) state, where the Cooper pairing takes place with nonzero momentum~\cite{Fulde:1964,Larkin:1965}, emerges in the phase diagram of an imbalanced 2D Fermi gas~\cite{Conduit:2008,Yin:2014,Sheehy:2015,Toniolo:2017}.
It must be worthwhile to elucidate how the antiparallel magnetic fields compete or cooperate with the FFLO state.

\acknowledgments
The authors thank Ippei Danshita and Shunsuke Furukawa for valuable discussions.
This work was supported by JSPS KAKENHI Grants No.~JP15K17727 and No.~JP15H05855, as well as International Research Center for Nanoscience and Quantum Physics, Tokyo Institute of Technology.

\end{document}